\documentstyle[epsfig]{aipproc}
\begin{document}

\title{UIT Astro-2 Observations\\ of NGC 4449}
\author{Robert S. Hill$^{*}$ \and Michael N. Fanelli$^{*}$ \and 
Denise A. Smith$^{\dagger}$ \and Ralph C. Bohlin$^{\ddagger }$ 
\and Susan G. Neff$^{\parallel }$ \and %
Robert W. O'Connell$^{\P }$ \and Morton S. Roberts$^{\S }$ \and Andrew M.
Smith$^{\parallel }$ \and Theodore P. Stecher$^{**}$}
\address{%
$^{*}$Hughes STX Corp., Code 681, NASA/GSFC, Greenbelt, MD 20771\\
$\dagger $NRC, Code 681, NASA/GSFC, Greenbelt, MD 20771\\
$^{\ddagger }$Space Telescope Science Insitute, 3700 San Martin Drive,
Baltimore, MD 21218\\
$^{\parallel }$Code 681, NASA/GSFC, Greenbelt, MD 20771\\
$^{\P }$University of Virginia, P. O. Box 3818, Charlottesville,
VA 22903\\
$^{\S }$National Radio Astronomy Observatory, Edgemont Road,
Charlottesville, VA 22903\\
$^{**}$Code 680, NASA/GSFC, Greenbelt, MD 20771\\
}
\maketitle

\begin{abstract} 
The bright Magellanic irregular galaxy NGC 4449 was observed by the
Ultraviolet Imaging Telescope (UIT) during the Astro-2 Spacelab mission in
March, 1995.  Far ultraviolet (FUV) images at a spatial resolution of $\sim
3$'' show bright star-forming knots that are consistent with the general
optical morphology of the galaxy and are often coincident with bright H II
regions.  Comparison of FUV with H$\alpha$ shows that in a few regions,
sequential star formation may have occurred over the last few Myr.  The bright
star forming complexes in NGC 4449 are superposed on a smooth, diffuse FUV
background that may be associated with the H$\alpha$ ``froth.''
\end{abstract}

\section*{Observations}

NGC 4449 is a bright Magellanic irregular galaxy at a distance of 5.4
Mpc\cite{rsh:bomans}. Its absolute FUV magnitude is estimated at $-20.2$, as
compared to $-18.8$ for the Large Magellanic Cloud (LMC), $-19.3$ for the Local
Group Sc spiral M33, and $-21.8$ for the SBc spiral M83\cite{rsh:rocket}. In
other words, NGC 4449 is comparable to conspicuous, star-forming spirals in FUV
luminosity, and it is brighter than the LMC by a factor of $\sim3.5$. This
signature of active, ongoing star formation is confirmed by H$\alpha $
imagery\cite{rsh:hg90}, which shows a large number of compact, bright H{\sc %
II} regions against a background of loops, filaments, and diffuse emission. The
southern bar is a region of particular complexity, with many bright near-point
sources and with bright filamentary structure seen in line emission.  

NGC 4449 was observed by the Ultraviolet Imaging Telescope (UIT\cite
{rsh:stecher92}) during the Astro-2 Space Shuttle mission in March, 1994. UIT
observations on the Astro-2 mission were in one of the FUV bands, usually
either B1 ($\lambda = 152$ nm, $\Delta\lambda = 35$ nm) or B5 ($\lambda = 162$
nm, $\Delta\lambda = 22$ nm). The two bandpasses are similar; B5 has a longer
short-wavelength cutoff to exclude dayglow.

\begin{figure} % fig 1
%\vspace{4in}
\centerline{\epsfig{file=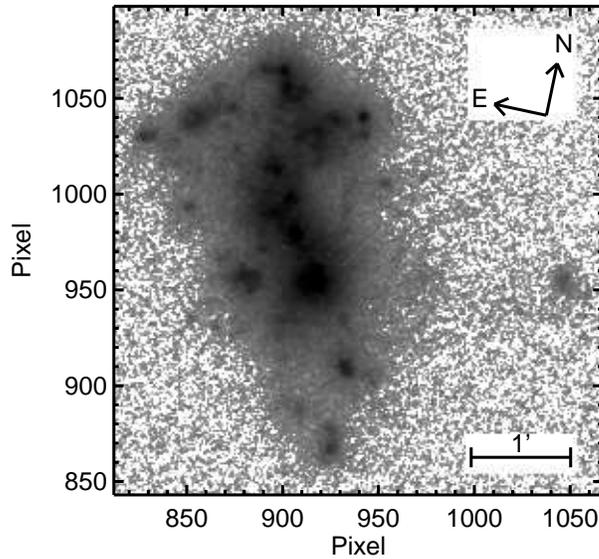}}
\caption{UIT FUV Image of NGC 4449 (B1 Filter).}
\label{rsh:fig1}
\end{figure}

Observations of NGC 4449 were obtained in both B1 and B5. In this paper, we
discuss the 900 s B1 image, which is the deepest one (Figure \ref{rsh:fig1}),
and the 500 s B5 image, which has the best spatial resolution.

NGC 4449 was also observed from the ground using the Goddard Fabry-Perot
Imager (GFPI). These observations, made by K.-P. Cheng, R. Oliversen, and P.
M. N. Hintzen, are discussed in an earlier paper based on FUV sounding
rocket imagery\cite{rsh:rocket}. The GFPI data consist of emission-line
images in H$\alpha $, H$\beta $, and nearby, narrow continuum bands.

As a result of analyzing the UIT data, the calibration of observations
reported in the sounding rocket paper\cite {rsh:rocket} has been revised.
In all bands, the reported fluxes are a factor of $\sim 2$ too high. 
Specifically, any single Balmer line flux should be multiplied by $0.6$, and
any single rocket UV flux should be multiplied by $\sim 0.5$.  The off-band
visual continuum measurements are affected in the same way as the line fluxes. 
In the UV case, the correction is less well defined because of a somewhat
uncertain bad-pixel factor, varying from source to source, which is included in
the rocket data, but is unnecessary for the UIT data.  In summary, the effect
on any ratio of two fluxes is $\sim 20$\%.  Therefore, the modeled source ages
reported in the rocket paper are not significantly affected.

\section*{Analysis}

The term {\em froth\/} has been coined to describe a certain morphology of
H$\alpha$-emitting material outside of typical H{\sc ii} regions.  Froth lacks
conspicuous embedded star clusters, and it is distinguished from simple diffuse
emission by a complex structure of bright filaments.  The froth is thought to
be the product of a mixed ionization mechanism, with energy supplied both by
ionizing stars and by shocks \cite{rsh:hg90}.

\begin{table}
\caption{Photometry of Selected Froth Regions}
\label{rsh:tab1}
\begin{tabular}{cccc}
\multicolumn{1}{c}{No.}&
$F_{H\alpha}$ & $f_{FUV}$ & Equivalent Age \\
& \multicolumn{1}{c}{(erg cm$^{-2}$s$^{-1}$)} &
\multicolumn{1}{c}{(erg cm$^{-2}$s$^{-1}$\mbox{\AA}$^{-1}$)} &
\multicolumn{1}{c}{(Myr)}\\
\tableline
1 & 2.33$\times 10^{-13}$ & 2.28$\times 10^{-14}$ & 4.0 \\ 
2 & 1.74$\times 10^{-13}$ & 1.81$\times 10^{-14}$ & 4.2 \\ 
3 & 1.00$\times 10^{-13}$ & 1.11$\times 10^{-14}$ & 4.4 \\ 
4 & 1.05$\times 10^{-13}$ & 1.33$\times 10^{-14}$ & 4.8 \\ 
\end{tabular}
\end{table}

A substantial contribution from photoionization is likely.  A plausibility
argument for this position can be made with a simple application of the FUV
image data.  We examine the ratio between H$\alpha$ surface brightness in any
given local area and the coincident FUV surface brightness.  If the ratio of
H$\alpha$ to FUV is consistent with that expected from an ionizing cluster,
then photoionization is a plausible mechanism.  Apertures $\sim 20$'' in
radius are superposed on 4 regions of diffuse, filamentary H$\alpha$ emission
and extended UV emission.  The H$\alpha $ and FUV fluxes for
these regions are given in Table \ref{rsh:tab1}, together with the implied
ages\cite{rsh:rocket}, without extinction corrections or background flux
subtraction.  The implied FUV surface brightnesses are equivalent to 1 or 2
unreddened early B stars per arcsec$^2$.  

\begin{figure} % fig 2
%\vspace{4in}
\centerline{\epsfig{file=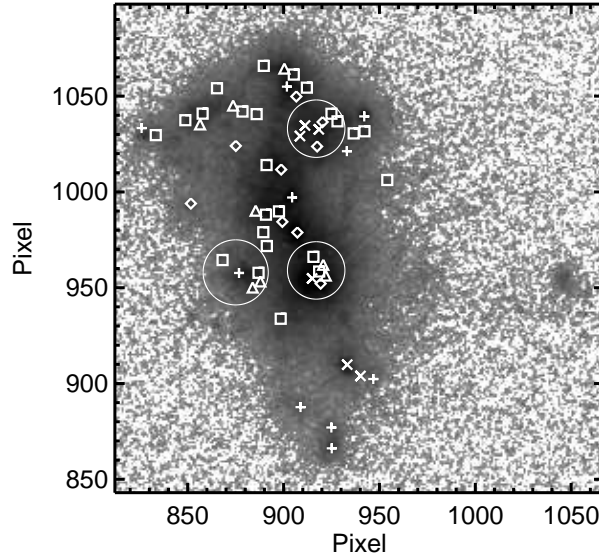}}
\caption{Compact H$\alpha$ + FUV Sources Plotted on B1 Image.  Symbols
indicate ages in Myr, as follows:  triangles, $<1$; squares, $1-3$; 
diamonds, $3-5$; plusses, $5-7$; crosses, $>7$.}
\label{rsh:fig2}
\end{figure}

In the case of compact, bright knots of H$\alpha $ and FUV emission, the age
computed from the ratio of the two fluxes has real meaning as the possible age
of a coeval cluster of stars.  Ages are computed from H$\alpha$ to FUV (B5)
flux ratios in circular apertures.  Sources are grouped by proximity and a
background surface brightness within  NGC 4449 is subtracted from each group.
Extinction is computed using the Balmer decrement, assuming that
$A_{FUV}/E(B-V) = 4.5$, which is similar to the Orion Nebula extinction
curve\cite{rsh:orion}.  Figure \ref{rsh:fig2} shows a map of the compact source
ages.  Three regions showing a spatial progression of ages are circled.  These
patterns may  indicate sequential star formation of the type discussed by
Elmegreen \& Lada\cite{rsh:elmelada,rsh:athome}.  In this scenario, the typical
spatial separation between stellar generations appears to be $\sim 100-300$ pc,
a factor of $\sim 5$ larger than in Elmegreen \& Lada's Galactic examples.


\begin{references}

\bibitem{rsh:bomans}  Bomans, D. J., Chu, Y.-H., \& Hopp, U. 1997, AJ, in press

\bibitem{rsh:rocket}  Hill, R. S., Home, A. T., Smith A. M., Bruhweiler, F.
C., Cheng, K.-P., Hintzen, P. M. N., \& Oliversen, R. J. 1994, ApJ, 430, 568

\bibitem{rsh:hg90}  Hunter, D. A. \& Gallagher, J. S. 1900, ApJ, 362, 480

\bibitem{rsh:stecher92}  Stecher, T. P. et al. 1992, ApJ, 395, L1

\bibitem{rsh:orion}  Bohlin, R. C. \& Savage, B. D. 1981, ApJ, 249, 109

\bibitem{rsh:elmelada}  Elmegreen, B. G. \& Lada, C. J. 1977, ApJ, 214, 725

\bibitem{rsh:athome}  Home, A. T., private communication

\end{references}
\end{document}